\documentclass[aps,prl,twocolumn,floatfix]{revtex4}
\usepackage{amsmath2000,bm, graphicx}

\newcommand{\beq}{\begin{equation}}
\newcommand{\eeq}{\end{equation}}
\newcommand{\bea}{\begin{eqnarray}}
\newcommand{\eea}{\end{eqnarray}}
\newcommand{\bml}{\begin{mathletters}}
\newcommand{\eml}{\end{mathletters}}

\begin{document}
\title{Exact Solution of the Mu\~noz-Eaton Model for Protein Folding}
\author{Pierpaolo Bruscolini}
\email{pbr@athena.polito.it}

\author{Alessandro Pelizzola}
\email{alex@athena.polito.it}
\affiliation{Dipartimento di Fisica \& INFM, Politecnico di Torino,
c.so Duca degli Abruzzi 24, I-10129 Torino}

\date{\today}

\begin{abstract}
A transfer-matrix formalism is introduced to evaluate
{\em exactly} the partition function of the  Mu\~noz-Eaton model, relating
the folding kinetics of proteins of known structure to their
thermodynamics and topology. This technique
can be used for a generic protein, for any choice of the energy and
entropy  parameters, and in principle allows the model to be used as a
first tool to characterize the dynamics of a protein of known native state
and equilibrium population. Applications to a $\beta$-hairpin and to
protein CI-2, with comparisons to previous results, are also shown.
\end{abstract}
\maketitle

Recent experimental findings on the folding of small proteins suggest
that, despite the complex microscopical dynamics of the proteins in
solution, the overall characteristics of the folding kinetics (e.g,
the equilibrium rates) could be quite simple, and principally related
to the topology of the native state. 
This has led to the construction of some simple models \cite{eatnat,
AlmBak99, GaFi99, Bak00, Amos} that, basing on the knowledge of the
native fold, aim to predict the kinetic properties and their changes
upon mutations. 

Even if,
usually, the interest  in  the folding  
is related to prediction of the native structure, 
the understanding of the folding process of proteins 
of known  structure
is nonetheless of great importance, both
for theoretical reasons, and 
because important diseases  are 
due 
to the aggregation of partially folded or misfolded proteins
\cite{landsbury99}. 

The importance of these models is that they are simple enough to be
dealt with quite easily, and have proved to provide relevant 
information on the folding
process  as a whole at physiological temperatures. 
They are useful to test the hypothesis  
that a reaction coordinate exists,
with one or a few major rate-limiting steps, and that motion
along it is ruled by the formation of just the native contacts, so
that all the others can be effectively averaged out.
So, in a sense, they are complementary 
of the more realistic all-atoms models, that, due to their 
computational cost, can be applied to characterize just a 
small part of the folding process at room temperature, or to follow 
the unfolding  at unrealistically high temperatures from a 
biological point of view.    

In this letter we focus on  the Mu\~noz-Eaton (ME) model \cite{eatnat,
eathair, eatprot}, where a protein is described by a chain of peptide bonds
that can live in just two states: native and unfolded. 
Two residues can interact only if they do so
in the native structure and if all the intervening residues in the
chain between them are native. 

In the application to several proteins \cite{eatprot}, the authors do
not consider a detailed dynamics, but rather relate the relaxation
rates to the free energy profiles $F_j$ (i.e. the free energy as a
function of the number of native bonds $j$).  Moreover, to reduce the
huge number of configurations that any protein can live in, even after
the discretization of the states, they calculate these profiles
resorting to the ``single/double/triple sequence'' approximation
(SSA/DSA/TSA): 
they consider 
only the contribution to the free energy of configurations where just
one/two/three stretches of native bonds are present, thus effectively
lowering the 
entropy of the unfolded state.
Exact solutions of the ME model under the assumption of 
homogeneous (i.e.\ residue independent) interactions 
have been carried out in \cite{Amos} for the
$\beta$-hairpin and the $\alpha$-helix structures,
 together with a mean field approximation whose  
accuracy, in comparison with a Monte Carlo simulation, was assessed
for protein CI-2.

The main result in this letter is the presentation of a way to drop
these approximations and calculate the exact free energy, correlation 
functions and  any other  relevant thermodynamical  quantity  for any
given protein, within the model assumptions. The existence  of  
an exact  solution  is interesting for at least two reasons: the  first,
is that, in general, exact solutions are rare, and provide useful 
benchmarks to test approximations and simulations. The second is that, 
in this case, the solution is also easy to implement, and allows 
the evaluation of a protein free energy in a few seconds of CPU time.
As an illustration, we apply our technique to the C-terminal 
$\beta$-hairpin of  streptococcal protein G B1 and to protein CI2, finding
relevant differences from approximate results.

In the ME model \cite{eatnat, eathair, eatprot} 
the state of a protein of $N+1$ residues 
can be described by a
binary variable $m_i$ for each peptide bond $i=1,\ldots,N$ (peptide
bond $i$ connects residues 
$i$ and $i+1$). The values $m_i=0, 1$ indicate that the bond is
unfolded or native, respectively. 
The  hamiltonian  (indeed a free energy function) reads
\beq
\! H(\{m_k\}) = \sum_{i=1}^{N-1} \sum_{j=i+1}^{N} \epsilon_{i  j} 
\, \Delta_{i j}
\prod_{k = i}^j m_k  - T \sum_{i=1}^{N} \Delta s_i \, m_i ~ , 
\label{Eq:H}
\eeq
The first term assigns an energy  $\epsilon_{i j} <0$ to the contact
between bonds $i$, $j$ (also accounting for the contact between
residues $i$, $j+1$). This energy gain is present 
provided that all the
bonds  from $i$ to $j$ are native, and that the contact 
exists also in the native structure 
($\Delta_{i j}=1$ in that case; $\Delta_{i j}=0$ otherwise).
The second term represents the  entropic cost 
$\Delta s_i < 0$ of ordering bond $i$ in the native state. 
From the above equation, it follows that the contribution 
$w(\{m_k\})\,  = \, \exp [- H(\{m_k\})/(R T) ]$ 
of any configuration
$\{m_k\}= (m_1, m_2, \ldots, m_N)$ to the partition function
${\cal Z} = \sum_{\{m_k\}} w(\{m_k\})$
is just the product of the weights of  the stretches of native bonds
contained in that configuration; namely:
\beq
\!w_{j, i} \!= \!\exp \!\left[- \frac{1}{R T} \left(\sum_{k=i}^{j-1}
\sum_{l=k+1}^{j} \!\!
\epsilon_{k  l} \, \Delta_{k  l} -T \sum_{k=i}^{j} \Delta s_k \right) 
\right], \!\!
\label{Eq:wji}
\eeq
for a native stretch going from bond $i$ to bond $j$
($i\le j$); for later use we define $w_{j \, j+1}=1$.

\begin{figure}
\includegraphics[width=45mm]{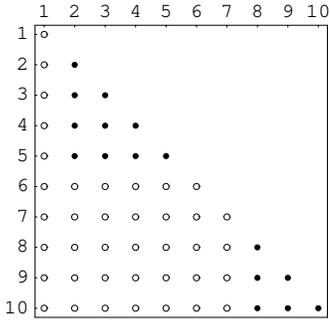}
\caption{2D description of a configuration 
(i.e $\{m_i\} \equiv \{m_{i, i}\}= 
(0,1,1,1,1,0,0,1,1,1)$) of the original model, for
$N=10$. Filled symbols represent $1$'s; empty symbols $0$'s. Notice
that, in any configuration, the $1$'s will group in  triangular regions
close to the diagonal, due to Eq.\ (\protect{\ref{Eq:mji}}). Hence, in
each row $j$, all the $1$'s will stay on the right, yielding just $j+1$
possible states for the row.  
In evaluating the transfer matrix, the
square angle vertex of the triangle, $(j,i)$ (here (5,2) and (10,8)),  
will be attributed the
weight $w_{j,i}$. } 
\label{Fig:oneconfig}
\end{figure}

We start
by mapping the original one-dimensional (1D) model, with long range
interactions, to a 2D one, with nearest-neighbour
interactions. We do so by introducing
\beq
m_{j, i} = \prod_{k = i}^j m_k~,
\label{Eq:mji}
\eeq
whence the obvious constraints $m_{j, i} = 0,~ 1$  and $m_{j, i}
= m_{j-1, i}\, m_{j, i+1}$ follow. The 2D representation of a
generic configuration of the original model is reported in
Fig.~\ref{Fig:oneconfig}.
It can be shown that, in the new variables,  it is possible to apply the
Cluster Variation Method \cite{cvm} to write a variational free energy
function, 
whose minimum provides the exact free energy of the model \cite{unpub}.
Here we follow an alternative route, and notice that
the states of row $j$ below 
the diagonal in the 2D model
are completely defined  specifying the number of 1's in the row.
Notice in fact that all the 1's stay necessarily at the right end of
the row (Fig.~\ref{Fig:oneconfig}):
this is the key observation, allowing the introduction of an efficient
transfer matrix formalism to solve the model.  In fact, while the
diagonal (i.e. the set of original variables of the model) can assume
$2^N$ configurations, row $j$ in the lattice (but the same is also
true for the columns) can assume only $j+1$ states, according to how
many 1's are present. Let $\bm{v}_k^j$ represent the state with 
$k$ 1's ($k=0, \ldots, j$) of row $j$. 
The transfer matrix from row $j+1$ to row $j$ will be defined by its
action on the vectors $\bm{v}_k^{j+1}$ by the equation:
\begin{subequations}
\label{Eq:Qmx}
\bea
Q_{j+1}^j (\lambda) \, \bm{v}_{k}^{j+1} &=&  \bm{v}_{k-1}^{j} ~,\;\; \text{for
$k=1, \ldots, j+1$;} \label{Eq:Qmx1}  \\
Q_{j+1}^j (\lambda) \, \bm{v}_{0}^{j+1} &=&  \sum_{k=0}^j \lambda^k ~
w_{j, j+1-k} ~ \bm{v}_{k}^{j} ~, 
\label{Eq:Qmx2}
\eea
\end{subequations}
where $\lambda$ is a dummy variable 
whose exponent takes into account 
the number of 1's (and hence, of native residues) being
introduced at row $j$. This will serve to generate the free energy
profiles at fixed number of native residues, as explained below.
In components, observing that, for each $j$, $\bm{v}_{k}^{j}$
($k=0,\ldots,j$) can be considered  the basis of a
$(j+1)$-dimensional space, we  write 
$(\bm{v}_{k}^{j})_l = \delta_{l, k+1}$ and
\beq
\left(Q_{j+1}^j (\lambda)\right)_{l,m} \!\!\!= (1- \delta_{m,1})\,
\delta_{l,m-1} + \delta_{m,1}\, \lambda^{l-1}\, w_{j, j+2-l}~,
\eeq
where $l=1,\ldots,j+1$ and $m=1,\ldots,j+2$.
Thus, $Q_{j+1}^j$ are rectangular matrices; the biggest one,
$Q_{N+1}^N$, is an $(N+1) \times (N+2)$ matrix.

The partition function of the model will be calculated as ${\cal Z} \equiv
{\cal Z}(\lambda=1)$, where
\beq
{\cal Z}({\lambda})  =  \bm{v}_{0}^{0}  \,\,
\Pi^0_{N+1} (\lambda) \,\, \bm{v}_{0}^{N+1}  =
 \sum_{j=0}^{N} Z_j ~ \lambda^j ~.
\label{Eq:Z}
\eeq
Here and below $\Pi^i_j (\lambda)\equiv Q^{i}_{i+1}(\lambda) \,\, \ldots
\,\, Q^{j-1}_{j} (\lambda)$.
From the above expressions it is clear that ${\cal Z}({\lambda})$ is the
generating function for the contributions $Z_j$ 
coming from the configurations
with fixed number of native bonds $j$: from here we recover the free
energy as a function of the native bonds as
\beq
F_j = - R \,T \log  Z_j~.
\label{Eq:Fj}
\eeq

In the same fashion it is possible to evaluate the average values of
$m_{j,i}$, Eq.~(\ref{Eq:mji}): 
introducing the matrix $P_{i,j}$: 
\beq
\begin{cases}
P_{i,j}\,  \bm{v}_{k}^{j} = 0 & \text{for $k< j-i+1$,}\\
P_{i,j}\,  \bm{v}_{k}^{j} = \bm{v}_{k}^{j} & \text{otherwise,}
\end{cases}
\label{Eq:Pmx}
\eeq 
we will have
\beq
\langle m_{j,i} \rangle = \frac{1}{ {\cal Z}} \bm{v}_{0}^{0} \,\,
\Pi^0_j(1) 
\,\, P_{i,j} \,\, 
\Pi^j_{N+1}(1) \,\,
\, \bm{v}_{0}^{N+1} \,. 
\label{Eq:mmedji}
\eeq
Since $m_{j,i}=1$ represents a native
stretch going from $i$ to $j$, $\langle m_{j,i} \rangle$ is the probability
of observing such a stretch.
Other important quantities in defining the folding pathways are the $\mu_{j,i}
= \langle (1-m_{i-1})
(\prod_{k=i}^j m_{k}) (1-m_{j+1}) \rangle$, 
which represent the
probability of a native stretch going from 
bonds $i$ to $j$,
preceded and followed by a non-native bond. They can be calculated as
\beq
\mu_{j,i} =  \frac{1}{ {\cal Z}} \bm{v}_{0}^{0} \,\,
\Pi^0_j \,\,
E_{i,j} \,\,  Q^{j}_{j+1} \,\, O_{j+1,j+1}\,\,
\Pi^{j+1}_{N+1}\,\,
 \bm{v}_{0}^{N+1}\,, 
\label{Eq:mu_ji}
\eeq
where $Q^{j-1}_{j} \equiv Q^{j-1}_{j} (\lambda=1)$; 
the matrix $E_{i,j}$ propagates only the state with a stretch of
$j-i+1$ native bonds between $i$ and $j$:
\beq
E_{i,j}\, \bm{v}_{k}^{j}\, =\,\delta_{k,j-i+1}\, \bm{v}_{j-i+1}^{j}\, ,
\label{Eij}
\eeq
while $O_{j,j}$ forces bond  $j$ to be non native:
\beq
O_{j,j} \,\bm{v}_{k}^{j}\, =\,\delta_{k,0} \,\bm{v}_{0}^{j}\, .
\label{Ojj}
\eeq

\begin{figure}
\includegraphics[width=50mm]{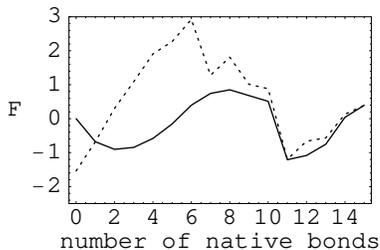}
\caption{Free energy profiles for the hairpin according to
Eq.~(\protect{\ref{Eq:Fj}}): x-axis: number of native bonds; y-axis:
contribution to free energy  (in Kcal/mol).
Exact solution (solid line); modified SSA 
(dotted line; see \protect{\cite{eathair}} for details).
Here $T=297$ K, $\Delta H_{hb}=-1.09$ Kcal/mol,
$\Delta G_{sc}=-2.03$ Kcal/mol, $\Delta S_{conf}=-3.12$ cal/(K mol), see text.
Notice that the native state minimum corresponds to just eleven native bonds:
the native state is such that the hydrophobic cluster is 
formed, but the terminal residues are disordered.
}
\label{Fig:hairpin}
\end{figure}

\begin{figure}
\includegraphics[width=45mm]{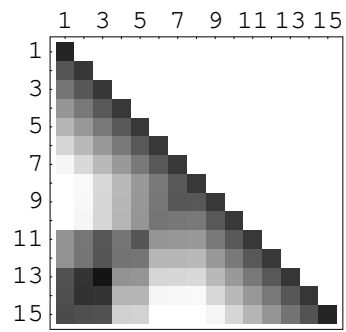}
\caption{Density plot of $- \log \mu_{j,i}$ (see
Eq.~(\protect{\ref{Eq:mu_ji}})) for the hairpin:
darker regions correspond to more probable stretches. It is possible to
recognize the  unfolding pathway as
$(13,3)-(12,3)-(11,3)-(11,4)-(11,5)-((10,5)\, {\text or}\, (11,6))-(10,6)-
((9,6)\, {\text or} \, (10,7))-(9,7)$, where the position $(j,i)$ in the matrix
corresponds to a native stretch from bond $i$ to $j>i$.   
}
\label{Fig:hairpinstretch}
\end{figure}

We have applied our technique to 
the 16 C-terminal  
residues from protein G B1, folding in a $\beta$-hairpin 
\cite{eatnat, eathair}. 
In order to compare our results with
those in \cite{eathair}, we
set  the contact matrix $\Delta_{i j}$ in Eq.\ (\ref{Eq:H}) according
to Fig.\ 1 in \cite{eathair}, and  specialize 
$\Delta s_i$  and $\epsilon_{i j}$  
to the values $\Delta s_i =\Delta S_{\text{{\it conf}}}$ and
$\epsilon_{i j} = p \Delta H_{hb} + q \Delta G_{sc}$, where $p$ is the
number of backbone-backbone hydrogen bonds between peptide bonds $i$
and $j$, and $q$ is the number of side-chain hydrophobic interactions
between residue $i$ and $j+1$; $\Delta H_{hb}$, $\Delta G_{sc}$ are
the (free) energy contributions of hydrogen bonds and ``hydrophobic'' 
interactions, respectively, while  $\Delta S_{conf}$ is the entropy
loss upon fixing a bond in its native conformation (refer to the
original article for a detailed discussion).
We fix 
$\Delta H_{hb}$, $\Delta G_{sc}$, $\Delta S_{conf}$ by fitting
$\langle m_{13,3} \rangle$,  
the population of the stretch with formed hydrophobic
cluster, to the experimental data in \cite{eathair}, Fig.~3. We find
that 
the exact solution 
Eq.\ (\ref{Eq:Fj}) yields  free energy profiles which are
smoother than those obtained with SSA, with considerably lower
barriers (see Fig.\ \ref{Fig:hairpin}). As expected, 
the differences are greater in
the unfolded part of the profile, where the entropic contribution of
partially ordered conformations dominates.
Even if we do not attempt a microscopic dynamics, upon  assuming that a 
single-residue dynamics in the equilibrium landscape may represent
the true kinetics we can try to characterize the (equilibrium) folding
barriers. To this end, we consider the values of the $\mu_{j,i}$ of
Eq.\ (\ref{Eq:mu_ji}).
They allow to follow the kinetic pathway of the hairpin,
since the length of the native stretches must increase on approaching
the native state, and it can be verified that, for the hairpin, it is
most likely that a $n$-bonds native stretch is created by adding a
native site at one end of $(n-1)$-stretch, rather than filling a 1-bond
gap  between two shorter stretches and merging them.
Fig.\ \ref{Fig:hairpinstretch} suggests  
that the barrier in the folding/unfolding
pathway corresponds to the formation of six-bond-long stretches,
either from peptide bond  5 (Y45-D46) to  10 (K50-T51) 
or from  6 (D46-D47) to 11 (T51-F52).
The presence of a folding barrier thus appears as a feature of the model,
and not just of SSA, and contrasts with the results in \cite{Dineal99}.
%
%

\begin{figure}
\includegraphics[width=60mm]{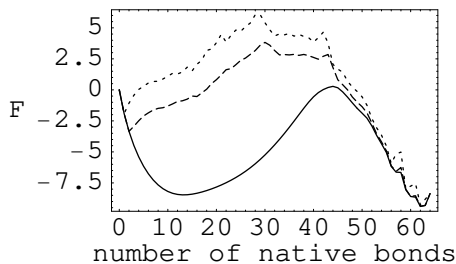}
\caption{Free energy profiles for protein CI-2 according to
Eq.~(\protect{\ref{Eq:Fj}}): x-axis: number of native bonds; y-axis:
contribution to free energy (in Kcal/mol). 
Exact solution (solid line); SSA
(dotted); DSA (dashed); see
\protect{\cite{eatnat}} 
for details.
Here $\epsilon= -0.550$ 
Kcal/mol,
$\Delta s_{0}= -1.327 $ 
cal/(K mol), 
$\Delta s_{1}= -3.863$ 
cal/(K mol), 
from the fit of calorimetric data: see text. 
}
\label{Fig:CI2Fj}
\end{figure}

Coming to proteins, we  
have applied our exact solution to the 65 terminal residues of protein
CI-2 (2CI2), choosing the values of $\epsilon_{i  j} \, \Delta_{i j}$ in
Eq.\ (\ref{Eq:H}) according to the following procedure.
As in \cite{eatprot}, we
define an atomic contact to be present if two nonhydrogen atoms, from
residues $i$ and $j>i+2$, are closer than 0.4 nm and  we  give an energy
$ k \epsilon$ to residues contacts involving $5 (k-1) < n_{at} \le 5
k$  atomic contacts. Moreover, we get the native secondary structure from
\cite{rcsbpdb} and assign to the peptide bond preceding a residue
marked by B, E, G, H, I, T, the entropy cost $\Delta s_1$ of the
``structured'' elements, while the other symbols will get  
the entropy $\Delta s_0$
of the ``less structured'' parts (i.e. coils, loops).          
Then, to fix $\epsilon$, $\Delta s_1$, $\Delta s_0$ 
we assume that the protein has a two-state thermodynamics:
the equilibrium population of the native state
will be given by $\theta = {\cal Z}_{nat}/{\cal Z} $, where 
${\cal Z}_{nat}$ is the sum of the contributions
from the right of the barrier in Fig.\ \ref{Fig:CI2Fj}.
Then, we fit the calorimetric data for $\Delta G =
G_u-G_n$ in \cite{Privalov}, 
by asking that $(1-\theta)/\theta = \exp
(-\Delta G/RT)$.
 
\begin{figure}
\includegraphics[width=50mm]{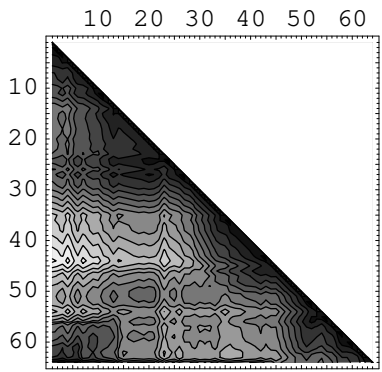}
\caption{Contour plot of $- \log \mu_{j,i}$ 
(Eq.~(\protect{\ref{Eq:mu_ji}})) for CI-2:
darker regions represent more likely stretches. 
Parameters as before. 
}
\label{Fig:CI2stretch}
\end{figure}

We can then study the differences
between the SSA/DSA and the exact free energy profiles, 
as well as the equilibrium values for the 
$\mu_{j,i}$ of  Eq.\ (\ref{Eq:mu_ji}). 
The results are reported in Figs. \ref{Fig:CI2Fj} and
\ref{Fig:CI2stretch} respectively, at the temperature of
equal native and unfolded population  $T=343.54$ K \cite{Privalov}.
%
It is clear that SSA and DSA deeply
underestimate the configurational entropy of the unfolded state,
yielding different barriers to folding and unfolding. 
The exact free
energy profile is also much smoother than the approximate ones.
Fig.\ \ref{Fig:CI2stretch} suggests that, in the unfolded state,
the  helix (13-24) and  the region including a turn (51-56) are the most
populated. Experimentally, the unfolded state is devoid of fixed
structure \cite{Fersht95}, but 
the N-terminal part of the helix, together with the
residues L50, I58 \cite{note}, 
constitute the folding nucleus: 
the model gives interesting clues
about this, even if 
with a turn overestimation \cite{note2}.  
The helix was already  noticed to be particularly populated in
\cite{AlmBak99, Amos, Mieal99}. 
%

From Fig.~\ref{Fig:CI2stretch} we can also speculate about the
folding pathway:
assuming 
an equilibrium folding dynamics, with the number of  native bonds 
increasing of at most one bond per time step, we try to characterize 
the unfolding pathway   by asking whether a $\mu_{j,i}$ will more likely drop 
an end,  producing   $\mu_{j-1,i}$ or $\mu_{j,i+1}$, or will split 
in two stretches $(i, k-1)$ and $(k+1,j)$  for some $k$. 
Upon reverting the unfolding pathway, the above analysis suggests 
that the folding  
starts with the formation of the turn at 
54-55; it grows up to the last residue and then towards the N-terminus; 
that folds last. 
Although this result correctly suggests that the helix alone 
cannot nucleate the folding \cite{Fersht95, AlmBak99}, 
this pathway does not agree with the
experimental one: this could be partially due to our one-residue
``dynamics''; notice though that the model cannot account for
interactions as those in the 
folding nucleus, 
involving residues that are not in the same  native
stretch: this is probably its  major limit.    
%

In conclusion, we have presented the complete exact solution for the 
thermodynamics of the ME model, in a form which is also easy 
to implement and to apply to real proteins. 
We have 
analyzed the cases of  a 
$\beta$-hairpin peptide and of protein CI-2, finding in both cases relevant 
differences   from previously published results. 
Our solution opens a wide range of applications to real proteins,
especially after a better characterization of the relationship between 
equilibrium properties and folding dynamics  is found:
work is in progress along these lines.

\begin{acknowledgments}
We are grateful to A. Maritan, F. Cecconi and A. Flammini for drawing our 
attention to the ME model, for communicating their results prior to
publication  and for  fruitful 
discussions. 
We are also indebted to M. Vendruscolo for kindly sharing with us  
his computer program to calculate contact maps. 
\end{acknowledgments}


\begin{thebibliography}{99}
\bibitem{note} In Ref. \protect{\cite{Fersht95}} Fersht studied the 64-terminal-residues fragment;
the residues' number here is shifted by one.
\bibitem{note2} Notice anyway that the greater stability of the turn,
and of the helix C-terminus over the N one, 
could be an artefact 
of dividing residues 
according to secondary structure:
the presence in those regions of ``less structured'' residues K25,
D53, D56 entails a smaller cost $\Delta s_0$ upon folding.
\end{thebibliography}
\end{document}